\documentclass[fleqn,10pt,twocolumn]{ICCAS2019}

\usepackage{graphicx}
\usepackage{caption}
\usepackage{subcaption}
\usepackage{mathptmx} % assumes new font selection scheme installed
\usepackage{times} % assumes new font selection scheme installed
\usepackage{amsmath} % assumes amsmath package installed
\usepackage{amssymb}  % assumes amsmath package installed
\begin{document}

\title{Noise Removal of FTIR Hyperspectral Images via MMSE}

\author{Chang Sik Lee${}^{1}$, Hyeong Geun Yu${}^{1}$, Dong Jo Park${}^{1}$, Dong Eui Chang${}^{1*}$, Hyunwoo Nam${}^{2}$, Byeong Hwang Park${}^{2}$}

\affils{ ${}^{1}$School of Electrical Engineering, Korea Advanced Institute of Science and Technology, \\
Daejeon, 34141, Korea (drancon@kaist.ac.kr, elloss@kaist.ac.kr, djpark@kaist.ac.kr, dechang@kaist.ac.kr)\\
${}^{*}$ Corresponding author \\
${}^{2}$CRB Defense Technology Directorate, Agency for Defense Development, \\
Daejeon, 34188, Korea (hyunwoonam@add.re.kr, pbh1348@add.re.kr)}

%\thanks{ \noindent
%   This paper is supported by my funding agencies.
%  }

\abstract{
    Fourier transform infrared (FTIR) hyperspectral imaging systems are deployed in various fields where  spectral information is exploited. Chemical warfare agent (CWA) detection is one of such fields and it requires a fast and accurate process from the measurement to the visualization of detection results, including noise removal. A general concern of existing noise removal algorithms is a trade-off between time and performance. This paper suggests a minimum mean square error (MMSE) approach as an efficient noise removal algorithm for FTIR hyperspectral images. The experimental result shows that the MMSE estimator spends less time to achieve comparable performance to the existing algorithms.  
}


\maketitle

%-----------------------------------------------------------------------

\section{Introduction}
The hyperspectral imaging (HSI) system is commonly used for obtaining information inaccessible with a general imaging system that utilizes only the visible light \cite{article:app1,article:app2,book:app3}. Especially, since several kinds of chemical warfare agents (CWAs) are detectable in the long-wave infrared (LWIR) spectral band, LWIR HSI devices have been studied for military and industrial purposes \cite{article:cwa1,article:cwa2}. A Fourier transform infrared (FTIR) HSI system is one general type of such equipment. The FTIR HSI system has the capability of capturing spectra over a wide range of spectral bands at remote distance \cite{article:ftir1}. Hence, the FTIR HSI system is employed for real-time detection and visualization of CWAs within a long range in the air \cite{article:ftir2,article:vis}.

In general, the process of detecting a CWA with the FTIR HSI system consists of multiple steps such as measuring a raw hyperspectral image, preprocessing the raw measurement, and detecting the CWA of interest.
Among these steps, the preprocessing plays a remarkable role in achieving good detection results and, in the preprocessing step, noise reduction accounts for removing undesired variance of signal, considered as noise. Numerous algorithms such as Gaussian filter \cite{article:gaussian} and the maximum noise fraction (MNF) transformation \cite{article:MNF} have been proposed for reducing random noises in FTIR hyperspectral images. However, those approaches suffer from a typical trade-off between time and performance \cite{article:tradeoff}, which means that, for better denoising results, we should use a time-consuming algorithm.

In this paper, we take a minimum mean square error (MMSE) approach \cite{book:mmse} to reduce  random noises in FTIR hyperspectral images. Through an experiment, it is shown that the  MMSE approach requires less computational resources than the MNF transformation, and the MMSE estimator improves the detection performance as much as the MNF transformation does.

\section{MMSE Estimator for Noise Reduction}

\subsection{Signal Model}
Given a hyperspectral image $I_{h}\in{\mathbb R}^{H\times W\times N}$, the measured spectrum ${\mathbf{z}}(x,y)\in {\mathbb{R}}^{1\times N}$ at pixel $(x,y)$ is defined as
\begin{equation}\label{eq:model}
	{\mathbf{z}}(x,y) = {\mathbf s}(x,y) + {\mathbf v}(x,y),
\end{equation}
where ${\mathbf s}(x,y)\in {\mathbb R}^{1\times N}$ is the signature spectrum at pixel $(x,y)$; $H$ and $W$ are the height and width of image $I_{h}$;  $N$ is the number of sampling points in the spectral range;  and $x$ and $y$ are integer pixel indices within the ranges of $[1,H]$ and $[1,W]$ respectively. The vector ${\mathbf v}(x,y)\in {\mathbb R}^{1 \times N}$ is Gaussian noise with the mean vector $\mathbf 0$ and the  covariance matrix $\sigma^{2}{\mathbf I}_{N}$. We assume that the standard deviation $\sigma$ of the noise vector ${\mathbf v}(x,y)$ is known.

Considering spatial correlation between a spectrum and its neighboring spectra, we can model the signature spectrum ${\mathbf s}(x,y)$ as a weighted sum of the true signature spectra ${\mathbf s}_{t}(p,q)$ at pixels $(p,q)$ in a set 
\begin{align*}
H= \left\{(p,q)\in {\mathbb N}^{1\times 2}| |p-x| \le \frac{k-1}{2}, |q-y| \le \frac{k-1}{2} \right\},
\end{align*}
where $k \times k$ is the window size of $H$ and $k$ is assumed to be a positive odd integer smaller than the number of points $N$ in the spectral range. Suppose a signature spectra matrix ${\mathbf S}(x,y)\in{\mathbb{R}}^{k^2 \times N}$ which is a collection of all the signature spectra in the window $H$. Then the matrix ${\mathbf S}(x,y)$ is modeled as
\begin{equation}\label{eq:model_spatial}
{\mathbf{S}}(x,y) = {\mathbf W}(x,y){\mathbf S}_{t}(x,y),
\end{equation}
where ${\mathbf S}_{t}(x,y)\in{\mathbb{R}}^{k^2 \times N}$ is a collection of all the true signature spectra in the window $H$ and ${\mathbf W}(x,y)\in {\mathbb{R}}^{k^2 \times k^2}$ is a weight matrix accounting for the effect of spatial correlation of neighboring spectra. We assume that the weight matrix ${\mathbf W}(x,y)$ is invertible and each of its row vectors sums up to 1.

\subsection{Minimum Mean Square Error Estimator}
Based on the model \eqref{eq:model}, an MMSE estimator $\hat{\mathbf s}(x,y)$ which minimizes the  cost
\begin{align*}
C(e)=\iint(\hat{\mathbf s}(x,y)-{\mathbf s}(x,y))^{2}f_{\mathbf S, Z}({\mathbf s}, {\mathbf z}) d{\mathbf s}d{\mathbf z}
\end{align*}
is obtained by
\begin{equation}\label{eq:MMSE_1}
\hat{\mathbf{s}}(x,y) = {\mathbf E}[{\mathbf s}(x,y)|{\mathbf z}(x,y)].
\end{equation}
Let the signature spectrum ${\mathbf s}(x,y)$ be a Gaussian random vector with the mean vector $\bar{\mathbf s}(x,y)$ and the covariance matrix $\Sigma_{\mathbf S}$ and be independent of the noise vector ${\mathbf v}(x,y)$. Then the measured spectrum ${\mathbf z}(x,y)$ is also a Gaussian random vector with the mean vector $\bar{\mathbf z}(x,y)$ and the covariance matrix $\Sigma_{\mathbf Z}$. The formula \eqref{eq:MMSE_1} of MMSE estimator $\hat{\mathbf s}(x,y)$ is computed as
\begin{equation}\label{eq:MMSE_2}
\hat{\mathbf{s}}(x,y) = \bar{\mathbf s}(x,y)+ ({\mathbf z}(x,y)-\bar{\mathbf z}(x,y)){\Sigma}^{-1}_{\mathbf Z}\Sigma_{\mathbf SZ},
\end{equation}
where $\Sigma_{\mathbf SZ}$ is the covariance matrix of random vectors ${\mathbf s}(x,y)$ and ${\mathbf z}(x,y)$; refer to \cite{book:mmse} for derivation of \eqref{eq:MMSE_1} and \eqref{eq:MMSE_2}.

Based on the assumption that the signature spectrum ${\mathbf s}(x,y)$ is independent of the noise signal ${\mathbf v}(x,y)$ and the noise signal ${\mathbf v}(x,y)$ has zero mean vector, we compute the covariance matrices $\Sigma_{\mathbf Z}$ and $\Sigma_{\mathbf SZ}$ as
\begin{align*}
\Sigma_{\mathbf Z} &= {\mathbf E}[({\mathbf s}'(x,y)+{\mathbf v}(x,y))^{T}({\mathbf s}'(x,y)+{\mathbf v}(x,y))] \\
&=\Sigma_{\mathbf S} + \sigma^{2}{\mathbf I}_{N}, \\
\Sigma_{\mathbf SZ} &={\mathbf E}[{\mathbf s}'^{T}(x,y)({\mathbf s}'(x,y)+{\mathbf v}(x,y))]\\
&=\Sigma_{\mathbf S} \\
&=\Sigma_{\mathbf Z} - \sigma^{2}{\mathbf I}_{N},
\end{align*}
where ${\mathbf s}'(x,y)={\mathbf s}(x,y)-\bar{\mathbf s}(x,y)={\mathbf s}(x,y)-\bar{\mathbf z}(x,y)$.
Thus, we can rewrite the formula \eqref{eq:MMSE_2} of MMSE estimator $\hat{\mathbf{s}}(x,y)$ as follows: 
\begin{equation}\label{eq:MMSE_4}
{\hat{\mathbf s}}(x,y) = \bar{\mathbf z}(x,y)+ ({\mathbf z}(x,y) - \bar{\mathbf z}(x,y)){\Sigma}_{\mathbf Z}^{-1}(\Sigma_{\mathbf Z} - \sigma^{2}{\mathbf I}_{N}).
\end{equation}

\subsection{Sample Covariance Matrix}
Assume that  source objects that contribute to the spectrum ${\mathbf z}(x,y)$ at pixel $(x,y)$ are invariantly detected entirely within its neighborhood $H$. By considering a set of measured spectra $\{{\mathbf z}(p,q) \mid (p,q)\in H\}$ as a sample set of the measured spectrum ${\mathbf z}(x,y)$ at pixel $(x,y)$, we can form a matrix ${\mathbf Z}(x,y) \in {\mathbb R}^{k^{2} \times N}$, which is a collection of the $k^{2}$ sample spectra, 
\begin{align}\label{eq:sample_matrix}
{\mathbf Z}(x,y)=[{\mathbf z}_{1}^{T}(x,y), {\mathbf z}_{2}^{T}(x,y), {\mathbf z}_{3}^{T}(x,y),\cdots, {\mathbf z}_{k^{2}}^{T}(x,y)]^{T},
\end{align}
where ${\mathbf z}_{i}(x,y)\in{\mathbb R}^{1\times N}, i \in \{1,2,3,\cdots,k^{2}\}$ denotes a sample spectrum in the sample set of ${\mathbf z}(x,y)$. 
Then the sample covariance matrix $\hat{\Sigma}_{\mathbf Z}$ at the pixel $(x,y)$ is computed as the following:
\begin{align}\label{eq:sample_cov}
\hat{\Sigma}_{\mathbf Z}&=\frac{1}{k^{2} - 1}{\mathbf Z}'^{T}(x,y){\mathbf Z}'(x,y),
\end{align}
where ${\mathbf Z}'(x,y) = {\mathbf Z}(x,y) - {\hat{\bar{\mathbf Z}}}(x,y)$ and ${\hat{\bar{\mathbf Z}}}(x,y)\in {\mathbb R}^{k^{2} \times N}$ includes $k^{2}$ duplications of the sample mean vector ${\hat{\bar{\mathbf z}}}(x,y)\in {\mathbb R}^{1 \times N}$ of ${\mathbf z}(x,y)$, which is given by
\begin{align}\label{eq:sample mean}
{\hat{\bar{\mathbf z}}}(x,y)&=\frac{1}{k^{2}}\sum_{i=1}^{k^{2}}{\mathbf z}_{i}(x,y).
\end{align}
Finally, replacing the covariance matrix $\Sigma_{\mathbf Z}$ in \eqref{eq:MMSE_4} by the sample covariance matrix $\hat{\Sigma}_{\mathbf Z}$ in \eqref{eq:sample_cov}, the MMSE estimator $\hat{{\mathbf{S}}}(x,y)\in {\mathbb R}^{k^{2}\times N}$ for the sample spectra ${\mathbf Z}(x,y)$ is given as
\begin{align}\label{eq:MMSE_final}
{\hat{\mathbf S}}(x,y) &= {\hat{\bar{\mathbf Z}}}(x,y) + {\mathbf Z}'(x,y)\hat{\Sigma}_{\mathbf Z}^{\#}(\hat{\Sigma}_{\mathbf Z} - \sigma^{2}{\mathbf I}_{N}),
\end{align}
where $\hat{\Sigma}_{\mathbf Z}^{\#}$ is the pseudo-inverse matrix of $\hat{\Sigma}_{\mathbf Z}$. The pseudo-inverse matrix $\hat{\Sigma}^{\#}_{\mathbf Z}$ is used instead of the inverse matrix $\hat{\Sigma}^{-1}_{\mathbf Z}$ because it is assumed that $k^{2} < N$ and the sample covariance matrix $\hat{\Sigma}_{\mathbf Z}$ does not have full rank.
We take the signature estimate ${\hat{\mathbf s}}(x,y)$ out of $\hat{{\mathbf{S}}}(x,y)$, which is the MMSE estimation corresponding to the spectrum ${\mathbf z}(x,y)$ measured at the pixel $(x,y)$.
%\begin{align}\label{eq:MMSE_final}
%{\hat{\mathbf s}}(x,y) &= {\hat{\bar{\mathbf z}}}(x,y) + ({\mathbf z}(x,y)-{\hat{\bar{\mathbf z}}}(x,y))\hat{\Sigma}_{\mathbf Z}^{-1}(\hat{\Sigma}_{\mathbf Z} - \sigma^{2}{\mathbf I}_{N}).
%\end{align}

\subsection{Dimension Reduction of MMSE computation}
By the singular value decomposition, the matrix ${\mathbf Z}'(x,y)$ is decomposed as
\begin{align}\label{eq:Z_SVD}
{\mathbf Z}'(x,y) &= U_{Z}D_{Z}V^{T}_{Z},
\end{align}
where $U_{Z}\in {\mathbb R}^{k^{2}\times k^{2}}$ and $V_{Z} \in {\mathbb R}^{N \times N}$ are orthogonal matrices, $D_{Z}\in {\mathbb R}^{k^{2} \times N}$ is a rectangular diagonal matrix that contains singular values of ${\mathbf Z}'(x,y)$ as its diagonal components. Then the sample covariance matrix $\hat{\Sigma}_{\mathbf Z}$ can be written as
\begin{align}\label{eq:Cov_SVD_1}
\hat{\Sigma}_{\mathbf Z} &= \frac{1}{k^{2} - 1}{\mathbf Z}'^{T}(x,y){\mathbf Z}'(x,y) \nonumber \\
&=\frac{1}{k^{2} - 1}V_{Z}D^{T}_{Z}D_{Z}V^{T}_{Z}.
\end{align}
Define an alternative sample covariance matrix $\tilde{\Sigma}_{\mathbf Z}\in {\mathbb R}^{k^{2} \times k^{2}}$ which is related with \eqref{eq:Z_SVD} as follows:
\begin{align}\label{eq:Cov_SVD_2}
\tilde{\Sigma}_{\mathbf Z} &= \frac{1}{k^{2} - 1}{\mathbf Z}'(x,y){\mathbf Z}'^{T}(x,y) \nonumber \\
&=\frac{1}{k^{2} - 1}U_{Z}D_{Z}D^{T}_{Z}U^{T}_{Z}.
\end{align}
Using \eqref{eq:MMSE_final}, \eqref{eq:Z_SVD}, \eqref{eq:Cov_SVD_1}, and \eqref{eq:Cov_SVD_2}, we can easily see that the following relation holds,
\begin{align}\label{eq:MMSE_reduction_eq}
{\hat{\mathbf S}}(x,y) &= {\hat{\bar{\mathbf Z}}}(x,y) + {\mathbf Z}'(x,y)\hat{\Sigma}_{\mathbf Z}^{\#}(\hat{\Sigma}_{\mathbf Z} - \sigma^{2}{\mathbf I}_{N}) \nonumber \\
&= {\hat{\bar{\mathbf Z}}}(x,y) +  (\tilde{\Sigma}_{\mathbf Z} - \sigma^{2}{\mathbf I}_{k^{2}})\tilde{\Sigma}_{\mathbf Z}^{\#}{\mathbf Z}'(x,y) \nonumber \\
&={\hat{\bar{\mathbf Z}}}(x,y) + U_{Z}(D_{Z}-(k^{2}-1)\sigma^{2}D^{\#}_{Z})V^{T}_{Z}.
\end{align}
The alternative sample covariance matrix $\tilde{\Sigma}_{\mathbf Z}$ allows the computation of \eqref{eq:MMSE_final} to be implemented in a more efficient way given by the second line of \eqref{eq:MMSE_reduction_eq}. Using \eqref{eq:model_spatial}, the final form of proposed noise reduction method is  
\begin{align}\label{eq:final_form}
\hat{\mathbf{s}}_{t}(x,y) &= {\mathbf \omega}_{c}(x,y)\{{\hat{\bar{\mathbf Z}}}(x,y) + (\tilde{\Sigma}_{\mathbf Z} - \sigma^{2}{\mathbf I}_{k^{2}})\tilde{\Sigma}_{\mathbf Z}^{\#}{\mathbf Z}'(x,y)\},
\end{align}
where ${\mathbf \omega}_{i}\in {\mathbb R}^{1 \times k^{2}}$ is $i$th row vector of the matrix ${\mathbf W}^{-1}(x,y)$, which is the inverse of weight matrix ${\mathbf W}(x,y)$. A positive integer $c$ is the row index corresponding to the center pixel of the window $H$.
\subsection{Computational Complexity}
In this section, we describe the computational complexity of MMSE estimator \eqref{eq:final_form} in terms of floating point operation (FLOP) \cite{article:FLOP} and compare the FLOP of MMSE with FLOP of other noise reduction algorithms, which are the Gaussian filter and the MNF transformation. The FLOP is counted by an open source tool named Counting the Floating Point Operations \cite{website:FLOP} with an additional rule for evaluating FLOP of eigendecomposition from \cite{article:LAPACK,article:eig1,book:eig2}.
\begin{table}
	\setlength{\extrarowheight}{1.5ex}
	\caption{FLOP counts of three noise reduction algorithms. The size of hyperspectral image is $H\times W\times N$ and the size of window is $k\times k$ where $k < H, k < W,$ and $k^{2}<N$.}
	\label{table:FLOP}
	\begin{center}
		\begin{tabu}to\linewidth{|X[0.7c,m]|X[1.2c,m]|}\hline
			Algorithm  \vspace{0.04 cm}&   FLOP \vspace{0.07 cm}\\\hline
			MMSE \vspace{0.04 cm}& $HW(4k^{4}N+6k^{6})$ \vspace{0.01 cm}\\\hline
			Gaussian filter \vspace{0.07 cm}& $HW2k^{2}N$ \vspace{0.07 cm}\\\hline
			MNF \vspace{0.04 cm}& $HW(4k^{4}N+17k^{6})$ \vspace{0.01 cm}\\\hline
		\end{tabu}
	\end{center}
\end{table}

Table \ref{table:FLOP} shows the FLOP of each algorithm with lower order terms ignored. As shown in the table, the Gaussian filter has the smallest FLOP count. Although the MMSE's FLOP count is larger than the FLOP count of Gaussian filter, it is still smaller than the FLOP count of MNF.
%The Gaussian filter has the FLOP of $HW2k^{2}N$ with the order of $k$ being 2 and it is the lowest order among the three algorithms' FLOP counts.
%Although the FLOP counts of MNF transformation and MMSE equally have  $4k^{4}N$ for their highest order terms, the coefficients of other terms are different. The MMSE's FLOP has $4k^6$, whereas the MNF transformation's FLOP has $17k^{6}$.
This is due to the fact that the computation of MNF transformation includes eigendecomposition whose FLOP is about $9n^{2}$ while the most complex computation in MMSE operation is matrix inversion whose FLOP is about $2n^{2}$ for a real symmetric matrix of size $n\times n$. Therefore, the MMSE approach consumes less computational resources than the MNF transformation as shown in Table \ref{table:FLOP}.
\begin{figure}
	\begin{center}
		\includegraphics[trim={0cm 0cm, 0cm, 0cm},clip,width = 0.3\textwidth,height = 0.15\textheight]{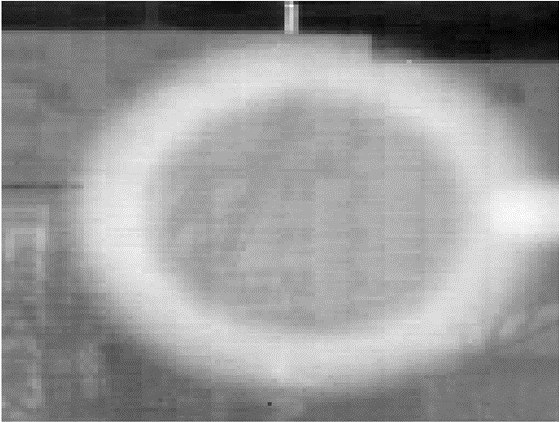}
		\caption{A hyperspectral image of gas cell.
			The gas cell is filled with sulfur hexaflouride ($\rm SF_{6}$) gas.}
		\label{fig:SF6}
	\end{center}
\end{figure}
\section{Experimental Result}
\begin{figure}
	\begin{subfigure}{.46\textwidth}
		\begin{center}
			\includegraphics[trim={0cm 0cm, 0cm, 0cm},clip,width = 1\textwidth,height = 0.3\textheight]{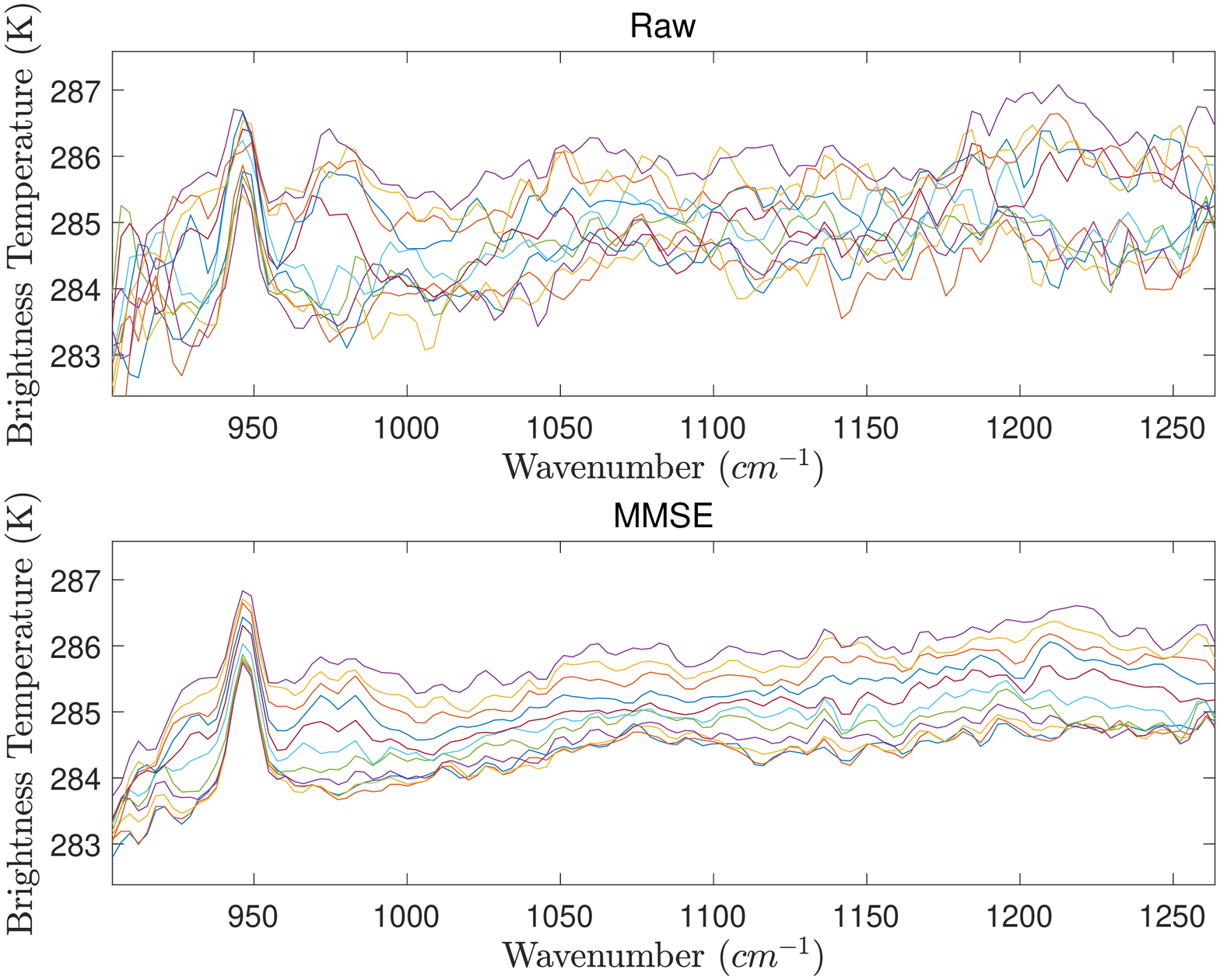}
			\caption{10 sample spectra of the raw image (upper) and the image denosied by MMSE (bottom)}
			\label{fig:Spectrum_Comparison_1}
		\end{center}
	\end{subfigure}
	\begin{subfigure}{.46\textwidth}
		\begin{center}
			\includegraphics[trim={0cm 0cm, 0cm, 0cm},clip,width = 1\textwidth,height = 0.3\textheight]{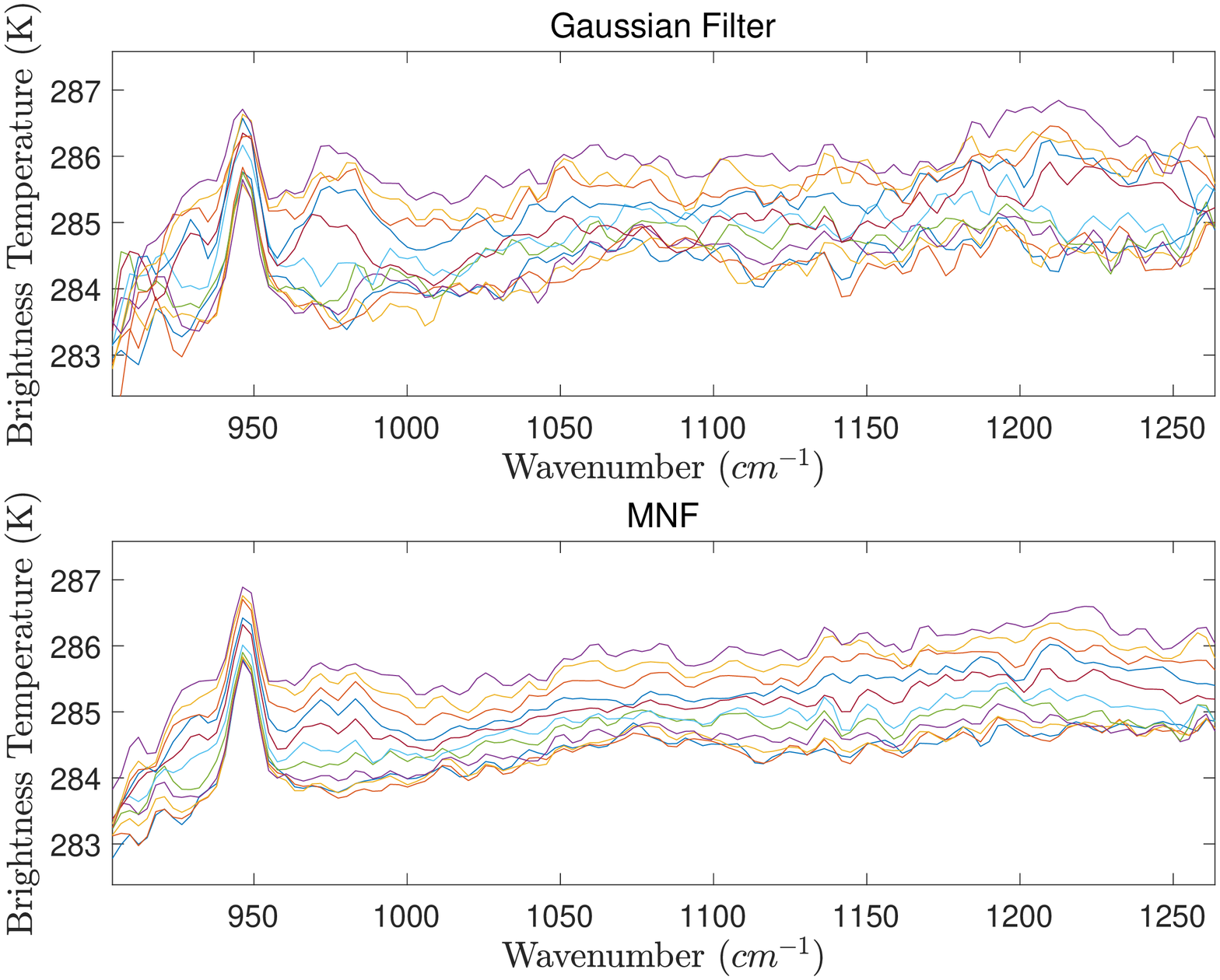}
			\caption{10 sample spectra of the image denoised by Gaussian filter (upper) and the image denosied by MNF (bottom)}
			\label{fig:Spectrum_Comparison_2}
		\end{center}
	\end{subfigure}
	\caption{10 sample spectra of the raw image and the images after noise reduction. The plots of noise reduced spectra largely show smaller variances than the raw spectra while the characteristic peaks of $SF_{6}$ at 950 $cm^{-1}$ are retained.}
	\label{fig:Spectrum_Comparison}
\end{figure}
\subsection{Experimental Setup}
In the experiment, a hyperspectral image is captured using a HI-90 manufactured by Bruker Optics, Germany. It is a remote FTIR hyperspectral imaging system with a spectral resolution of $3.2\enspace cm^{-1}$ in the spectral range of $900 \sim 1260\enspace cm^{-1}$ and a spatial resolution of $128 \times 128$ pixels.

We use a hyperspectral image $I_{h}\in{\mathbb R}^{128\times 128\times 128}$of a gas cell containing sulfur hexaflouride ($\rm SF_6$) with background of a building's wall for the experiment, as shown in Fig. \ref{fig:SF6}. The noise reduction performance of the MMSE estimator is given by the result of $\rm SF_6$ detection established by adaptive subspace detector (ASD) \cite{article:ASD}, which is shown by receiver operation characteristics (ROC) curve. Then the ASD ROC curve of MMSE estimator is compared with other cases: ASD ROC curves of the raw image, the image denoised by Gaussian filter, and the image denoised by MNF transformation. The window size $k$ is set to $3$ and the noise standard deviation $\sigma$ is assumed as $\sqrt{0.9}$. The weight vector $\omega_{c}(x,y)$ is a vectorized Gaussian weight centered at the pixel $(x,y)$ with standard deviation of $1$. Those algorithms are implemented and applied with Matlab R2017a.
%\begin{figure}
%\begin{center}
%\includegraphics[trim={0cm 0cm, 0cm, 0cm},clip,width = 0.3\textwidth,height = 0.15\textheight]{HI90.png}
%%\includegraphics[width=6.5cm]{test.eps}
%\caption{A picture of HI-90, Bruker Optics, Germany.}
%\label{fig:HI90}
%\end{center}
%\end{figure}
\begin{figure}
	\begin{center}
		\includegraphics[trim={0cm 0cm, 0cm, 0cm},clip,width = 0.45\textwidth,height = 0.27\textheight]{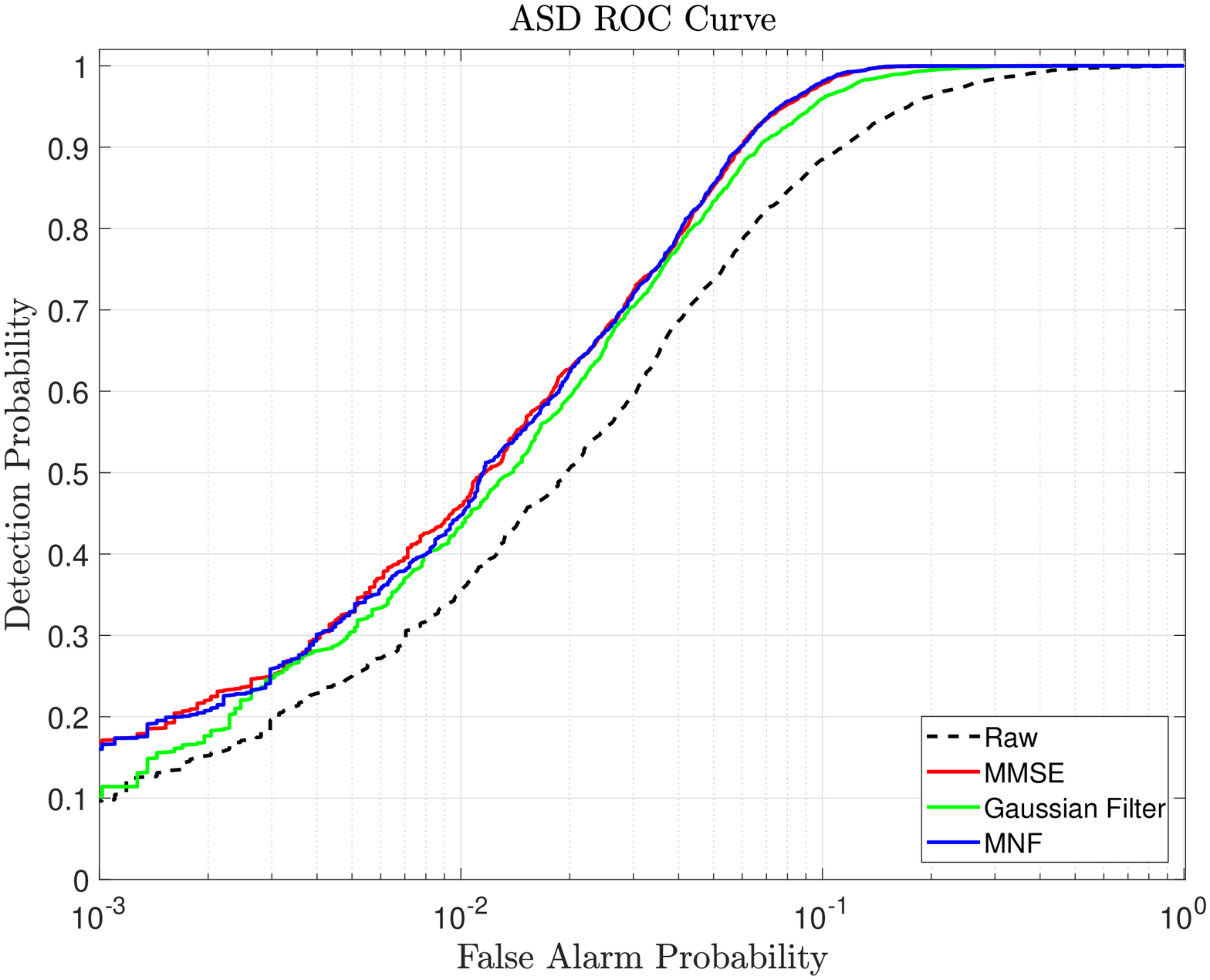}
		\caption{ROC curves of ASD for the raw image (raw, dotted black line), the image denoised by MMSE estimator (MMSE, red line), the image denoised by Gaussian filter (Gaussian Filter, green line), and the image denoised by MNF transformation (MNF, blue line). The MMSE's detection probability is higher than those of the raw image and the Gaussian filter, and is comparable to the MNF's ROC curve}
		\label{fig:ROC}
	\end{center}
\end{figure}
\subsection{Performance Comparison}
The results are shown in Figs. \ref{fig:Spectrum_Comparison} and \ref{fig:ROC}. Fig. \ref{fig:Spectrum_Comparison} displays 10 sample spectra of the cases specified in Section 3.1. At the top, raw measurements are shown and the subsequent plots are spectra denoised by MMSE, Gaussian filter, and MNF. Compared to the raw spectra in the upper plot of Fig. \ref{fig:Spectrum_Comparison_1}, the spectra in the other plots in Fig. \ref{fig:Spectrum_Comparison} show less variances overall, while preserving the characteristic peak of the target gas $SF_{6}$  around the wavenumber of 950 $cm^{-1}$. This suggests that the effect of Gaussian noise is reduced as consequences of the noise reduction algorithms.

Figure \ref{fig:ROC} shows the ROC curves of ASD for the cases tested in the experiment. Compared to the ROC curve of raw image, the MMSE estimator improves the ASD's detection performance, which means that the MMSE estimator \eqref{eq:final_form} successfully reduces the effect of Gaussian noise vector ${\mathbf v}(x,y)$ so that the ASD can effectively detect the desired gas spectra. Moreover, in the aspect of ASD's detection performance, the MMSE estimator makes larger enhancement than Gaussian filter does and the ROC curve enhanced by MMSE is on a par with the MNF transformation's result.
\begin{table}
	\setlength{\extrarowheight}{0.7ex}
	\caption{Computational resources for the noise reduction algorithms. The upper row shows the computing time and the bottom row shows the FLOP count for each algorithm.}
	\label{table:time}
	\begin{center}
		\begin{tabu}to\linewidth{|X[1.5c]|X[c]|X[c]|X[c]|}\hline
			&   \vspace{0.01 cm} MMSE   &  Gaussian filter \vspace{0.15 cm}  & \vspace{0.01 cm} MNF  \\\hline
			time (s) & 1.40 & 0.30 & 2.43   \\\hline
			FLOP ($\times 10^{9}$) & 1.02 & 0.03 & 1.15  \\\hline
		\end{tabu}
	\end{center}
\end{table}
\subsection{Efficiency Comparison}
To compare the computational load for running the noise reduction algorithms, we have measured the computation time and FLOP counts taken to execute each algorithm in the experiment.
%Since the computing time can vary in large range depending on the computing power of hardwares, for a definite comparison, we also have counted FLOPs required for the algorithms.
As shown in Table \ref{table:time}, the MMSE approach requires $42.39 \%$ less computing time and $11.30 \%$ less FLOP count than the MNF approach. The experimental result complies with the conclusion of Section 2.4 and it indicates that MMSE is a more efficient noise reduction algorithm than the MNF transformation.
\section{Conclusion}
This paper proposes applying an MMSE to the task of removing random noises from a real FTIR hypersepctral image. Based on the general assumption \cite{article:app2,book:app3,article:tradeoff} that objects captured in a pixel are invariant within its adjacent pixels, we compute the sample covariance matrix of measured spectrum to facilitate the MMSE estimator. The experimental result shows that the MMSE is an effective and time-efficient noise reduction algorithm for a real FTIR hyperspectral image by comparing the performance and computation load of MMSE estimator to the other noise reduction algorithms. For further research, deep learning based approaches \cite{book:nn} recently proposed for denoising images will be applied to noise removal of FTIR hyperspectral images.

\section{Acknowledgements}
This work was supported by the Agency for Defense Development of the Republic of Korea.

%%%%%%%%%%%%%%%%% BIBLIOGRAPHY IN THE LaTeX file !!!!! %%%%%%%%%%%%%%%%%%%%%%
%%---------------------------------------------------------------------------%%
%

%
%%--------------------------------------------------------------------%%

\end{document}